\documentclass[aps,pra,english,  notitlepage,superscriptaddress,showpacs,nofootinbib]{revtex4-1}
\usepackage{mathptmx}
\usepackage[T1]{fontenc}
\usepackage[latin9]{inputenc}
\usepackage{color}
\usepackage{babel}
\usepackage{textcomp}
\usepackage{bm}
\usepackage{graphicx}
\usepackage{amsmath}
\usepackage{amsthm}
\usepackage{amssymb}
\usepackage{epstopdf}
\usepackage[unicode=true,pdfusetitle,
bookmarks=true,bookmarksnumbered=false,bookmarksopen=false,
breaklinks=true,pdfborder={0 0 0},pdfborderstyle={},backref=false,colorlinks=true]
{hyperref}
\hypersetup{
	urlcolor=blue, citecolor=blue, linkcolor=blue}

\makeatletter
\theoremstyle{plain}
\newtheorem{thm}{\protect\theoremname}
\theoremstyle{plain}

\theoremstyle{definition}
\newtheorem{example}{\protect\examplename}
\theoremstyle{plain}
\newtheorem{lemma}{\protect\lemmaname}

\makeatother

\providecommand{\examplename}{Example}
\providecommand{\propositionname}{Proposition}
\providecommand{\theoremname}{Theorem}
\providecommand{\lemmaname}{Lemma}

\begin{document}
	
	\title{Genuine  hidden nonlocality without entanglement: from the perspective  of local discrimination }
	
	\author{Mao-Sheng Li}
	\email{li.maosheng.math@gmail.com}
	\affiliation{ School of Mathematics,
		South China University of Technology, Guangzhou
		510641,  China}

	\author{Zhu-Jun Zheng}\email{zhengzj@scut.edu.cn}
	\affiliation{ School of Mathematics,
		South China University of Technology, Guangzhou
		510641,  China}

	\begin{abstract}
		
		Quantum nonlocality without entanglement  is a fantastic phenomenon in quantum theory. This kind of quantum nonlocality is based on the task of    local discrimination of quantum states. Recently,  Bandyopadhyay and Halder [\href{https://journals.aps.org/pra/abstract/10.1103/PhysRevA.104.L050201}{Phys. Rev. A \textbf{104}, L050201 (2021)}]  studied the problem: is there any set of orthogonal  states which can be locally distinguishable, but under some   orthogonality preserving local measurement, each outcome will lead  to a locally indistinguishable set. We say that the set with such property has hidden nonlocality.  Moreover, if such phenomenon can not arise from discarding subsystems which is termed as local irredundancy, we call it   genuine hidden nonlocality. There, they presented several   sets of  entangled states with genuine hidden nonlocality. However, they doubted the existence of  a set without entanglement but  with genuine hidden nonlocality.   In this paper, we eliminate this doubt by constructing a series of sets without entanglement but whose nonlocality can be genuinely activated.
		We derive a method to tackle with the local irredundancy problem  which is a key tricky for the systems whose local dimensions are composite numbers. Unexpectedly, the constructions of genuine  hidden nonlocal sets without entanglement seems to be easier than that with entanglement. Therefore, this kind of nonlocality is rather different from the Bell nonlocality.

	\end{abstract}
	\maketitle
	
	\section{Introduction}
	The predictions of quantum theory
	are incompatible with those  based on the concept of locality \citep{Bell-1964,Brunner-et-al-2014}. Such phenomenon is known as quantum nonlocality. 	Quantum nonlocality is usually revealed by violating Bell type inequalities \citep{CHSH-1969,Freedman-Causer-1972,Aspect+-1981,Aspect+1982,Hensen+2015,handsteiner+2017,Rosenfeld+,BIG-BELL} which can only arise from entangled states.  The nonlocality arising from this way is termed as Bell nonlocality \citep{Brunner-et-al-2014,Bell-1964,CHSH-1969}.  It is well-known that entanglement \cite{Entanglement-horodecki} can not be generated  from separable states under local operations. However, there exists some states which admit a local hidden variable model (hence is local) but violating Bell type inequalities  after   judicious local filters are applied.  This kind of states is said to have hidden nonlocality \cite{Popescu-1995,Gisin-1996,navascues+-2011,Palazuelos-2012,Klobus+2012,Hirsch+2013}.

	Except the Bell nonlocality, there are other forms of nonlocality that have attracted our attention. In fact, the local indistinguishability  of some    set of orthogonal quantum states has been widely used to illustrate the phenomenon of quantum nonlocality.    In this setting, two or more observers share the parts of a  composite quantum system prepared in a state from a known orthogonal set.  The  task is to identify the state by performing   local operations and classical communication (LOCC). Note that we can always identify the state correctly if global measurements are allowed as the set of states are assumed to be orthogonal.  If such task can be  accomplished perfectly under LOCC, we say that the set is \emph{locally distinguishable},
	otherwise, \emph{locally indistinguishable}.
	Bennett et al. \cite{ben99} presented the first example 
	of  orthogonal product states that are locally indistinguishable   and   named such a phenomenon as  quantum nonlocality without entanglement.   The nonlocality here is in the sense that there exists some quantum information that could be inferred from global measurement but cannot be read from local correlations of the subsystems.   Since then,  this kind of nonlocality    has been studied extensively (see Refs. \cite{ben99,ben99u,Walgate-2000, Ghosh-2001,grois01,walgate-2002,divin03,HSSH,Ghosh-2004,rin04,Watrous-2005,fan-2005,Nathanson-2005,Wootters-2006,Hayashi-etal-2006,nis06,Duan2007,Duan-2009,feng09,Bandyo-2011,BGK-2011,Yu-Duan-2012,BN-2013,Cosentino-2013, Yus15,  B-IQC-2015,childs13,Cosentino-Russo-2014, Ha21,B-IQC-2015,  Li15, Yang13,zhang14,wang15,zhang15,Yang15,xu16-1,Xu-16-2,zhang16,zhang16-1,Xu-17,Wang-2017-Qinfoprocess,Zhang-Oh-2017,zhang17-1,halder,  Li20,  Li18,Halder1909,Xu20b, Xu20a, Halder20c}  for an
	incomplete list).   Most studies are focused on identifying sets of  orthogonal product states or sets of the maximally entangled states. In addition, a stronger manifestation of nonlocality  in multipartite systems has been studied which is  based on the notion of local irreducibility in all biparitions of subsystems \cite{Halder19,Zhang1906,Shi20S,Tian20,Wang21,Shi21,Shi21b,Banik21}.  An orthogonal set of pure states in multipartite quantum system is called \emph{locally irreducible} if it is not
			possible to eliminate one or more states from the set by orthogonality-preserving local measurements. 
	
	The local indistinguishability of  quantum states has   been practically applied in quantum cryptography primitives such as   data hiding \cite{Terhal01,DiVincenzo02} and secret sharing \cite{Markham08,Rahaman15,WangJ17}. Therefore, local indistinguishability of  quantum states can be considered as a  resource in  quantum  information processing.  If there are only  locally distinguishable sets   at hand, how can we transfer them into   resources that have applications in data hiding?  This is what   Bandyopadhyay and Halder  \cite{Bandyopadhyay21} recently studied. In fact, they studied the problem:  is there any set of orthogonal  states which can be locally distinguishable, but under some   orthogonality preserving local measurement, each outcome will lead  to a locally indistinguishable set. As there are some trivial sets with this property,  they introduced the concept of local irredundancy. An orthogonal set is said  to be   locally redundant if it remains
	orthogonal after discarding one or more subsystems. Otherwise, it is called to be locally irredundant.  
	If a locally irredundant set satisfies the aforementioned property, then we said its nonlocality can be activated  genuinely. There, they give several   examples of such sets with entanglement. However, deeper research  on this property  remains to be explored. For example, is there any locally distinguishable   set without entanglement whose nonlocality  can be genuinely activated?  Which  state spaces  have sets with this property? In this paper, we tend to  solve the two problems partially.

	The rest of this article is organized as follows. In Sec. \ref{sec:Pre}, we review the concept of genuine activation of nonlocality (genuine hidden nonlocality).  Then  we study the constructions
of sets with genuine hidden nonlocality with respect to two
cases: this first is cardinality preserving while the second is
cardinality decreasing.   In Sec. \ref{sec:Stable}, we present   examples for the first case and discuss its  constructions in multipartite systems.     In Sec. \ref{sec:Decreasing},  we give  a example for the second  case and also  and discuss its constructions in multipartite systems.        Finally, we draw a conclusion  in     section \ref{sec:Con}.

	\section{Preliminary}\label{sec:Pre}

	It is well known that a set of quantum states is perfectly distinguished by global measurement if and only if the given set of states are mutually orthogonal. Therefore, every  LOCC protocol that distinguishes a set of orthogonal
	states is a sequence of orthogonality-preserving-local-measurements (OPLM).	 In this paper, we  consider this specific class of LOCC measurements, i.e., OPLM.	
	
	Now we give a brief review of the main problem studied in Ref. \cite{Bandyopadhyay21}.	  Suppose  that  $\mathcal{S}$ is  an orthogonal set of multipartite states which is locally distinguishable.  The participants aim to find some OPLM $\mathbb{M}$ such that for each outcome $\mu$ of this measurement, they end up to  a new orthogonal set $\mathcal{S}_{\mu}^{\prime}$ which is locally indistinguishable.  If we could find such measurement,  we call that the nonlocality of $\mathcal{S}$  can be  \emph{activated} (Compared with setting in the Bell nonlocality , we  also call that $\mathcal{S}$  has \emph{hidden nonlocality} here).  However, the authors in Ref. \cite{Bandyopadhyay21} have pointed out that there are some trivial cases  which  arise from taking partial trace. Such cases would happen if the original set is local redundancy, i.e., the set remains orthogonal if we discard one or more subsystems. If the set $\mathcal{S}$ is locally irredundant and satisfies the aforementioned property, we can call that the nonlocality of $\mathcal{S}$ can be \emph{genuinely activated} (And we also call that $\mathcal{S}$  has \emph{genuine hidden nonlocality} here).  Note that the cardinality of  $\mathcal{S}_{\mu}^{\prime}$
	always satisfies $\left|\mathcal{S}_{\mu}^{\prime}\right|\leq\left|\mathcal{S}\right|.$ According to the differences of the cardinality, we separate the genuine hidden nonlocality into two types.
	\begin{enumerate}
		\item[{\rm(a)}]{\bf	Genuine hidden nonlocality of type I:} $\mathcal{S}$ is    locally distinguishable and locally irredundant. And there is  some OPLM $\mathbb{M}$ such that for each outcome $\mu$ of this measurement, the set of  post-measurement states $\mathcal{S}_{\mu}^{\prime}$  is locally indistinguishable and $|\mathcal{S}_{\mu}^{\prime}|=|\mathcal{S}|$.	 Moreover, if  for each outcome $\mu$ of this measurement, the set of  post-measurement states $\mathcal{S}_{\mu}^{\prime}$  is locally irreducible, we call it has genuine hidden strong form nonlocality of type I.
		\item[{\rm(b)}] {\bf	Genuine hidden nonlocality of type II:} $\mathcal{S}$ is    locally distinguishable and locally irredundant. And  there is some OPLM $\mathbb{M}$ such that for each outcome $\mu$ of this measurement, the set of  post-measurement states $\mathcal{S}_{\mu}^{\prime}$  is locally indistinguishable. In addition, there is some outcome $\mu$ such that    $|\mathcal{S}_{\mu}^{\prime}|<|\mathcal{S}|. $	 
	\end{enumerate}

	In this paper, we will study how to construct sets  without entanglement but with genuine hidden nonlocality of type I or II. Throughout this paper, we will use the following notations. Let $d\geq 2$ be an integer. Considering a  quantum system $\mathcal{H} $ of dimension $d$, we usually denote $\{|0\rangle, |1\rangle, \cdots, |d-1\rangle \}$ as its a computational basis. We also use the notation $|i+j\rangle$ (w.r.t. $|i-j\rangle$) which presents  $|i\rangle+|j\rangle$ (w.r.t. $|i\rangle-|j\rangle$ ). Moreover, we use $|+_n\rangle\equiv \sum_{i=0}^n|i\rangle,$ and  $|{}_m+_n\rangle\equiv \sum_{i=m}^n|i\rangle$ where $m<n$.  We also use $\mathbb{I}_{d}$ represent the identity operator on this system, i.e., $\mathbb{I}_{d}=\sum_{i=0}^{d-1}|i\rangle \langle i|.$   And the states throughout this paper may be unnormalized.

	\section{Genuine hidden nonlocality of type I}\label{sec:Stable}
	
	In this section, we study how to construct sets without entanglement but have genuine hidden nonlocality of type I. We will use the local  indistinguishability of  the  sets  without entanglement as follows.
	
	\begin{figure}[h]
		\centering
		\includegraphics[width=0.49\textwidth,height=0.3\textwidth]{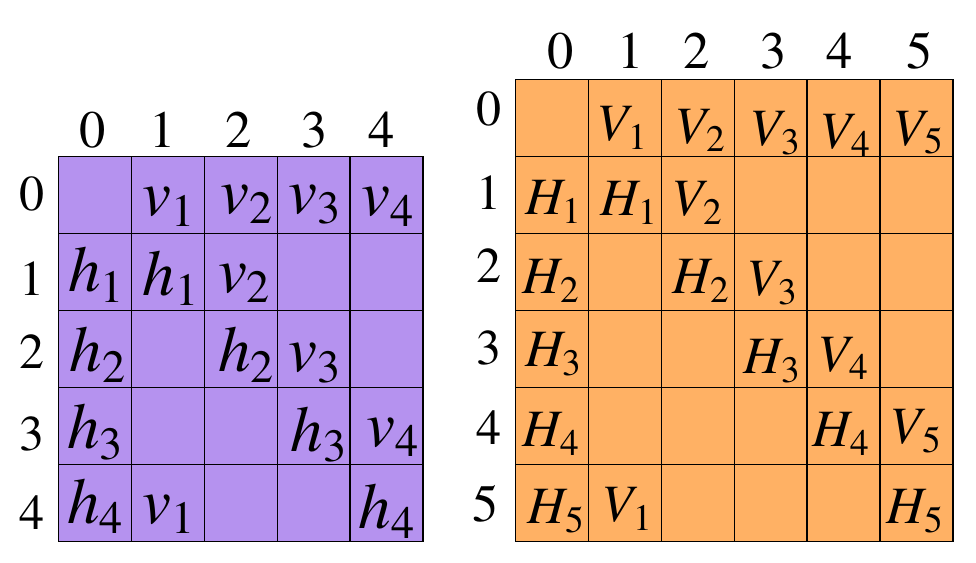}
		\caption{This shows two examples of the  structures of  the product states constructed in Theorem \ref{thm:UPB_Nonlocality} in $\mathbb{C}^5\otimes \mathbb{C}^5$ and $\mathbb{C}^6\otimes \mathbb{C}^6$. The squares indicated by the same label represent  a  unique state.  For examples, there are two squares (that is, $(2,0),$ and $(2,2)$) with label `$h_2$' in the left hand side graph, they correspond to the state $|2\rangle|0-2\rangle\in \mathbb{C}^5\otimes \mathbb{C}^5$  and  there are two squares (that is, $  (0,5),$ and $(4,5)$) with label `$V_5$' in the right hand side graph, they correspond to the state $|0-4\rangle|5\rangle\in \mathbb{C}^6\otimes \mathbb{C}^6.$  }\label{fig:thm1_Yu}
	\end{figure}

	\begin{thm}[\cite{Yus15}]\label{thm:UPB_Nonlocality}
		In $\mathbb{C}^d\otimes \mathbb{C}^d$ system with $d\geq 3$, the following $2d-1$  orthogonal pure product states are locally indistinguishable:
		$$\{|n\rangle|\delta_n\rangle\}_{n=1}^{d-1}\cup \{|\delta_n\rangle|n_+\rangle\}_{n=1}^{d-1}\cup\{|+_{(d-1)}\rangle|+_{(d-1)}\rangle\}$$
		where $|\delta_n\rangle\equiv |0-n\rangle$ and $n_+=n+1$ for $1\leq n\leq d-2$ while $n_+=1$ for $n=d-1$.
	\end{thm}
	
	Now let's start with  the following example, the construction of whose states is inspired from the above Theorem for the cases $\mathbb{C}^5\otimes \mathbb{C}^5$ and $\mathbb{C}^6\otimes \mathbb{C}^6$ (see Fig. \ref{fig:thm1_Yu} for their structures).

	\begin{figure}[h]
	\centering
	\includegraphics[width=0.48\textwidth,height=0.48\textwidth]{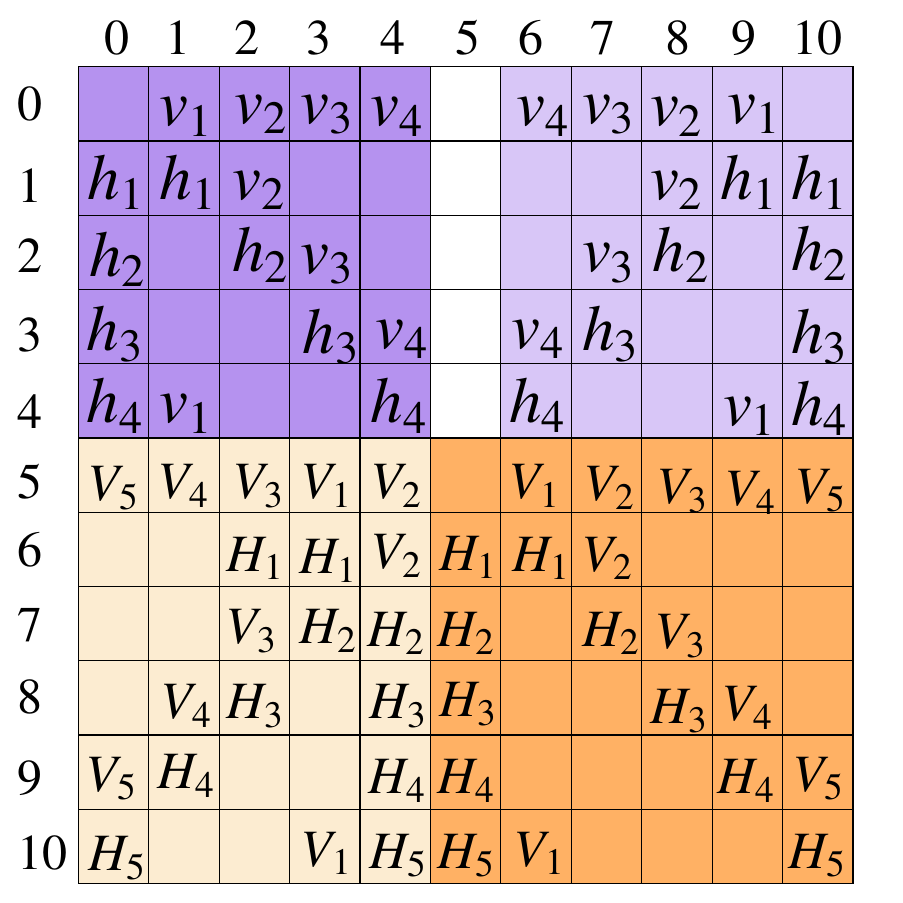}
	\caption{This shows the  states structure of set of product states in Example  \ref{ex-gan11} .  The squares indicated by the same label represent  a  unique state. For examples, there are four squares (that is, $(1,0), (1,1),(1,9),$ and $(1,10)$) with label `$h_1$', they correspond to the state $|\psi_1\rangle:=|1\rangle|0-1+9-10\rangle,$  and  there are four squares (that is, $  (5,2),(5,8),$ $(7,2)$, and $(7,8)$) with label `$V_3$', they correspond to the state $|\phi_8\rangle:=|5-7\rangle|8-2\rangle.$  The first five rows are axisymmetric about the middle column, and the first two columns and the last two columns of the last six rows are also axisymmetric about the middle column. }\label{fig:example1_11}
\end{figure}
	
	\begin{example}
		\label{ex-gan11} 
		Let $\mathcal{S}$ be the set of product states in $\mathbb{C}^{11}\otimes\mathbb{C}^{11}$ whose elements are listed as below (see Fig. \ref{fig:example1_11})):
		\begin{equation*}	
			\begin{array}{rclrcl}
				\left|\psi_{1}\right\rangle  & \equiv & |1\rangle|0- 1 +9-10\rangle,& 	\left|\psi_{5}\right\rangle  & \equiv & |0-4\rangle|1-9\rangle,\\ \left|\psi_{2}\right\rangle  & \equiv & |2\rangle|0-2 +8-10\rangle, & \left|\psi_{6}\right\rangle  & \equiv &| 0-1\rangle|2-8\rangle,\\
				\left|\psi_{3}\right\rangle  & \equiv & |3\rangle|0-3+7-10\rangle,&  \left|\psi_{7}\right\rangle  & \equiv & |0-2\rangle|3-7\rangle,\\	\left|\psi_{4}\right\rangle  & \equiv &| 4\rangle|0-4+6-10\rangle,
				& 	\left|\psi_{8}\right\rangle  & \equiv &| 0-3\rangle|4-6\rangle,\\
				\left|\phi_{1}\right\rangle  & \equiv & |6\rangle|5-6+2-3\rangle,& 	\left|\phi_{6}\right\rangle  & \equiv & |5-10\rangle|6-3\rangle,\\ \left|\phi_{2}\right\rangle  & \equiv & |7\rangle|5-7+3-4\rangle, & \left|\phi_{7}\right\rangle  & \equiv &| 5-6\rangle|7-4\rangle,\\
				\left|\phi_{3}\right\rangle  & \equiv & |8\rangle|5-8+2-4\rangle,&  \left|\phi_{8}\right\rangle  & \equiv & |5-7\rangle|8-2\rangle,\\	\left|\phi_{4}\right\rangle  & \equiv &| 9\rangle|5-9+1-4\rangle,
				& 	\left|\phi_{9}\right\rangle  & \equiv &| 5-8\rangle|9-1\rangle,	\\
				\left|\phi_{5}\right\rangle  & \equiv &| 10\rangle|5-10+0-4\rangle,
				& 	\left|\phi_{10}\right\rangle  & \equiv &| 5-9\rangle|10-0\rangle,				
			\end{array}
		\end{equation*}
		and $|\psi_{9}\rangle\equiv |+_4\rangle|+_{10}\rangle,$ and  $|\phi_{11}\rangle\equiv |{}_5+_{10}\rangle|+_{10}\rangle.$		  
		Then the set $\mathcal{S}$ has genuine hidden nonlocality of type I. 
	\end{example}

	Now we show that $\mathcal{S}$ is locally distinguishable.    For each $|\Theta\rangle=|\theta_1\rangle|\theta_2\rangle  \in \mathcal{S}$, we denote 
	$$
	\mathcal{C}(|\Theta\rangle):=\{(i,j)\in \mathbb{Z}_{11}\times \mathbb{Z}_{11} \big |   \langle i|  \theta_1\rangle   \langle j  |\theta_2\rangle  \neq 0\}.$$  We call it the coordinates of the state $|\Theta\rangle$. Note that except the states $|\psi_9\rangle$ and $|\phi_{11}\rangle$, the coordinates of other states in $\mathcal{S}$  are  all of cardinality $4$ and these states can be formed into two types: their coordinates lie on the same line and are labeled with $h_i$ or $H_i$ in Fig. \ref{fig:example1_11} (For examples,  $\mathcal{C}(|\psi_1\rangle)=\{(1,0),(1,1),(1,9),(1,10)\}$) and their coordinates  lie on the four corners of a  rectangle  and are labeled with $v_j$ or $V_j$ in Fig. \ref{fig:example1_11} (For example,    $\mathcal{C}(|\phi_6\rangle)=\{(5,3),(5,6),(10,3),(10,6)\}$). Moreover,   for different states $|\Theta_1\rangle,|\Theta_2\rangle\in \mathcal{S}\setminus \{|\psi_9\rangle, |\phi_{11}\rangle\}$, their coordinates are disjoint, i.e., $\mathcal{C}(|\Theta_1\rangle)\cap \mathcal{C}(|\Theta_2\rangle)=\emptyset.$
	 
	First, Alice perform the measurement  $\mathbb{M}_1^A:=\{\pi_i^A= |i\rangle\langle i |\  \big |  \ i\in \mathbb{Z}_{11}\}.$ 
 If the outcome of $\mathbb{M}_1^A$ is `$i$', by     the construction of the set $\mathcal{S}$ and  the above observation,   the postmeasurement states of  Bob's part     are mutually orthogonal and hence can be distinguished. For example, if the   outcome of $\mathbb{M}_1^A$ is `$5$',  the state must be one of $\{|\phi_{7,8,9,10,11}\rangle\} $  whose Bob's part are the following five orthognal states $
|7-4\rangle, |8-2\rangle, |9-1\rangle,|10-0\rangle, |+_{10}\rangle$ respectively.

	Now we prove that the set $\mathcal{S}$ satisfies the second property. Suppose Bob perform the measurement
	$\mathbb{M}^B:=\{\pi_1^B= \sum_{i=0}^4|i\rangle\langle i|,\ \pi_2^B=  \sum_{j=5}^{10}|j\rangle\langle j|\}.$
	
	If the outcome of $\mathbb{M}^B$ is `1', the states are transferred 
	to 	
	\begin{equation*}	
		\begin{array}{rclrcl}
			\left|\tilde{\psi}_{1}\right\rangle  & = & |1\rangle|0- 1\rangle,& 	\left|\tilde{\psi}_{5}\right\rangle  & = & |0-4\rangle|1\rangle,\\ \left|\tilde{\psi}_{2}\right\rangle  & = & |2\rangle|0-2 \rangle, & \left|\tilde{\psi}_{6}\right\rangle  & = &| 0-1\rangle|2\rangle,\\
			\left|\tilde{\psi}_{3}\right\rangle  & = & |3\rangle|0-3\rangle,&  \left|\tilde{\psi}_{7}\right\rangle  & = & |0-2\rangle|3\rangle,\\	\left|\tilde{\psi}_{4}\right\rangle  & = &| 4\rangle|0-4\rangle,
			& 	\left|\tilde{\psi}_{8}\right\rangle  & = &| 0-3\rangle|4\rangle,\\
			\left|\tilde{\phi}_{1}\right\rangle  & = & |6\rangle|2-3\rangle,& 	\left|\tilde{\phi}_{6}\right\rangle  & = & |5-10\rangle|3\rangle,\\ \left|\tilde{\phi}_{2}\right\rangle  & = & |7\rangle|3-4\rangle, & \left|\tilde{\phi}_{7}\right\rangle  & = &| 5-6\rangle|4\rangle,\\
			\left|\tilde{\phi}_{3}\right\rangle  & = & |8\rangle|2-4\rangle,&  \left|\tilde{\phi}_{8}\right\rangle  & = & |5-7\rangle|2\rangle,\\	\left|\tilde{\phi}_{4}\right\rangle  & = &| 9\rangle|1-4\rangle,
			& 	\left|\tilde{\phi}_{9}\right\rangle  & = &| 5-8\rangle|1\rangle,	\\
			\left|\tilde{\phi}_{5}\right\rangle  & = &| 10\rangle|0-4\rangle,
			& 	\left|\tilde{\phi}_{10}\right\rangle  & = &| 5-9\rangle|0\rangle,	 \\
			\left|\tilde{\psi}_{9}\right\rangle  & = &| +_4\rangle|+_4\rangle,
			& 	\left|\tilde{\phi}_{11}\right\rangle  & = &| {}_5+_{10}\rangle|+_4\rangle,						
		\end{array}
	\end{equation*}
	which are mutually orthogonal and  contain $9$ states $\{|\tilde{\psi}_i\rangle\}_{i=1}^9$ that  is known to be  locally indistinguishable in $\mathbb{C}^5\otimes  \mathbb{C}^5$. 
	
	If the outcome of $\mathbb{M}^B$ is `2', the states are transferred 
	to 	
	\begin{equation*}	
		\begin{array}{rclrcl}
			\left|\tilde{\psi}_{1}\right\rangle  & = & |1\rangle|9-10\rangle,& 	\left|\tilde{\psi}_{5}\right\rangle  & = & |0-4\rangle|9\rangle,\\ \left|\tilde{\psi}_{2}\right\rangle  & = & |2\rangle|8-10\rangle, & \left|\tilde{\psi}_{6}\right\rangle  &= &| 0-1\rangle|8\rangle,\\
			\left|\tilde{\psi}_{3}\right\rangle  & = & |3\rangle|7-10\rangle,&  \left|\tilde{\psi}_{7}\right\rangle  & = & |0-2\rangle|7\rangle,\\	\left|\tilde{\psi}_{4}\right\rangle  & = &| 4\rangle|6-10\rangle,
			& 	\left|\tilde{\psi}_{8}\right\rangle  & =&| 0-3\rangle|6\rangle,\\
			\left|\tilde{\phi}_{1}\right\rangle  & = & |6\rangle|5-6\rangle,& 	\left|\tilde{\phi}_{6}\right\rangle  & = & |5-10\rangle|6\rangle,\\ \left|\tilde{\phi}_{2}\right\rangle  & = & |7\rangle|5-7\rangle, & \left|\tilde{\phi}_{7}\right\rangle  & = &| 5-6\rangle|7\rangle,\\
			\left|\tilde{\phi}_{3}\right\rangle  & = & |8\rangle|5-8\rangle,&  \left|\tilde{\phi}_{8}\right\rangle  & = & |5-7\rangle|8\rangle,\\	\left|\tilde{\phi}_{4}\right\rangle  & = &| 9\rangle|5-9\rangle,
			& 	\left|\tilde{\phi}_{9}\right\rangle  & =&| 5-8\rangle|9\rangle,	\\
			\left|\tilde{\phi}_{5}\right\rangle  & = &| 10\rangle|5-10\rangle,
			& 	\left|\tilde{\phi}_{10}\right\rangle  & = &| 5-9\rangle|10\rangle, \\
			\left|\tilde{\psi}_{9}\right\rangle  & = &| +_4\rangle|{}_5+_{10}\rangle,
			& 	\left|\tilde{\phi}_{11}\right\rangle  & = &| {}_5+_{10}\rangle|{}_5+_{10}\rangle,					
		\end{array}
	\end{equation*}
	which are mutually orthogonal and  contain  $11$ states $\{|\tilde{\phi}_i\rangle\}_{i=1}^{11}$  that  is known to be  locally indistinguishable in $\mathbb{C}^6\otimes  \mathbb{C}^6$. \qed
	
	\vskip 10pt
	
	The above construction and  argument can be easily extend to $\mathbb{C}^{d}\otimes \mathbb{C}^{d}$ for any prime number $d\geq 11$.  Here we give one more example without proof.

		\begin{figure}[h]
		\centering
		\includegraphics[width=0.49\textwidth,height=0.49\textwidth]{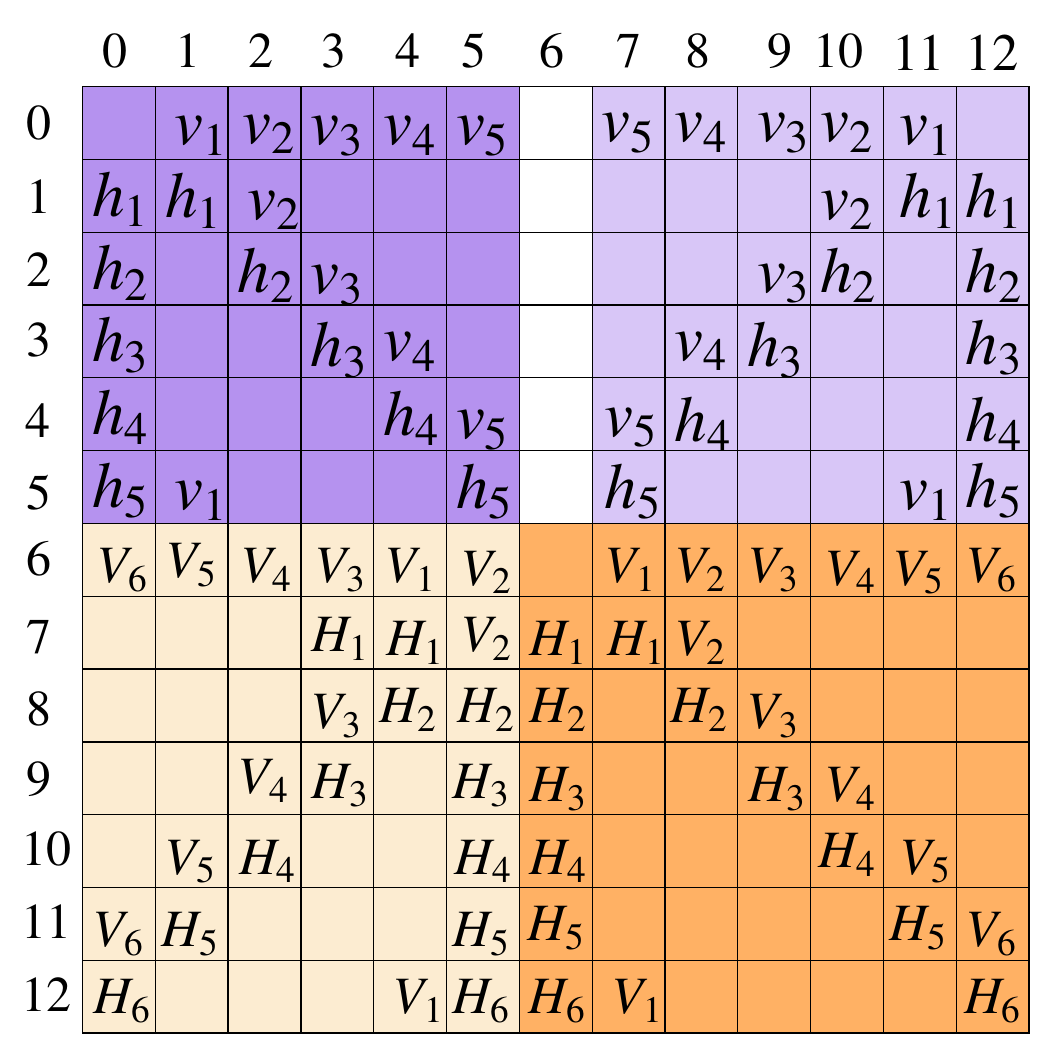}
		\caption{This shows the  states structure of set of product states in Example  \ref{ex-gan13} .   The first six rows are axisymmetric about the middle column, and the first three columns and the last three columns of the last seven rows are also axisymmetric about the middle column. }\label{fig:example1_13}
	\end{figure}
	
	\begin{example}
		\label{ex-gan13} 
		Let $\mathcal{S}$ be the set of product states in $\mathbb{C}^{13}\otimes\mathbb{C}^{13}$ whose elements are listed as below (see Fig. \ref{fig:example1_13})):
		\begin{equation*}	
			\begin{array}{rclrcl}
				\left|\psi_{1}\right\rangle  & \equiv & |1\rangle|0- 1 +11-12\rangle,& 	\left|\psi_{6}\right\rangle  & \equiv & |0-5\rangle|1-11\rangle,\\ \left|\psi_{2}\right\rangle  & \equiv & |2\rangle|0-2 +10-12\rangle, & \left|\psi_{7}\right\rangle  & \equiv &| 0-1\rangle|2-10\rangle,\\
				\left|\psi_{3}\right\rangle  & \equiv & |3\rangle|0-3+9-12\rangle,&  \left|\psi_{8}\right\rangle  & \equiv & |0-2\rangle|3-9\rangle,\\	\left|\psi_{4}\right\rangle  & \equiv &| 4\rangle|0-4+8-12\rangle,
				& 	\left|\psi_{9}\right\rangle  & \equiv &| 0-3\rangle|4-8\rangle,\\
				\left|\psi_{5}\right\rangle  & \equiv &| 5\rangle|0-5+7-12\rangle,
				& 	\left|\psi_{10}\right\rangle  & \equiv &| 0-4\rangle|5-7\rangle,\\
				\left|\phi_{1}\right\rangle  & \equiv & |7\rangle|6-7+3-4\rangle,& 	\left|\phi_{7}\right\rangle  & \equiv & |6-12\rangle|7-4\rangle,\\ \left|\phi_{2}\right\rangle  & \equiv & |8\rangle|6-8+4-5\rangle, & \left|\phi_{8}\right\rangle  & \equiv &| 6-7\rangle|8-5\rangle,\\
				\left|\phi_{3}\right\rangle  & \equiv & |9\rangle|6-9+3-5\rangle,&  \left|\phi_{9}\right\rangle  & \equiv & |6-8\rangle|9-3\rangle,\\	\left|\phi_{4}\right\rangle  & \equiv &| 10\rangle|6-10+2-5\rangle,
				& 	\left|\phi_{10}\right\rangle  & \equiv &| 6-9\rangle|10-2\rangle,	\\
				\left|\phi_{5}\right\rangle  & \equiv &| 11\rangle|6-11+1-5\rangle,
				& 	\left|\phi_{11}\right\rangle  & \equiv &| 6-10\rangle|11-1\rangle,			\\	
				\left|\phi_{6}\right\rangle  & \equiv &| 12\rangle|6-12+0-5\rangle,
				& 	\left|\phi_{12}\right\rangle  & \equiv &| 6-11\rangle|12-0\rangle,
			\end{array}
		\end{equation*}
		and $|\psi_{11}\rangle\equiv |+_5\rangle|+_{12}\rangle,$ and  $|\phi_{13}\rangle\equiv |{}_6+_{12}\rangle|+_{12}\rangle.$		  
		Then the set $\mathcal{S}$ has genuine hidden nonlocality of type I. 
	\end{example}

	For further discussion in the composite dimension, we need the following lemma.

	\begin{lemma}\label{lemma:nonorthogonal}
		Let $\mathcal{N}$ be a quantum channel from  system $\mathcal{H}_A$ to     system $\mathcal{H}_B$.  Then $\mathcal{N}$ preserves non-orthogonality. That is,  for any   density operators $\rho,\sigma$ of systems $\mathcal{H}_A$,   $$\langle \rho,\sigma\rangle_A\neq 0,\  \Rightarrow  \ \langle \mathcal{N}(\rho),\mathcal{N}(\sigma)\rangle_B\neq 0.$$
		Here, the inner product $ \langle \rho,\sigma\rangle_A:=\mathrm{Tr}[\rho^\dagger \sigma].$ Particularly, if we choose $\mathcal{N}$ to be the partial trace operation, then we have that  the  partial trace operation preserves non-orthogonality. 
	\end{lemma}
	Lemma  \ref{lemma:nonorthogonal} can be understood as a corollary of the well known result: a set of quantum states is perfect distinguishable under global measurement if and only if the given set of states are mutually orthogonal. For the sake of completeness,  we    give a direct proof  in the APPENDIX.

	\vskip 5pt

\noindent	{\bf Method for checking local irredundancy:}
Now we return to discuss the genuine  activation of nonlocality in general  dimension. Suppose  $d=2k+1\geq 11$, we can construct the set $\mathcal{S}$ to  be the union of $\{|\psi_i\rangle\}_{i=1}^{2k-1}$ and   $\{|\phi_i\rangle\}_{j=1}^{2k+1}$ similarly to Example  \ref{ex-gan11} and  Example  \ref{ex-gan13}.  In the following, we show that even for the case of composite number $d$,  the above set of states is   locally  irredundant whenever $d\geq 11$.	 Let $d=p_1p_2\cdots p_n$, where $p_i|d$ is some prime divisor of $d$ (here we assume that $3\leq p_1\leq p_2\leq\cdots\leq p_n$ and $n\geq2$). Suppose that $\mathcal{H}_X=\mathbb{C}^d$, it can be factored into an $n$-particles systems $\otimes_{i=1}^n\mathcal{H}_{X_i}$  where  $\mathcal{H}_{X_i}=\mathbb{C}^{p_i}$ and $X\in\{A,B\}$.  In fact, we will show that Alice and Bob can not  preserve the orthogonality of the given states by   discarding one or more of their   subsystems. Written  $|\psi_{i}\rangle=|\psi_i\rangle_A|\psi_i\rangle_B$ for $i=1,2,\cdots,2(k-1)$, one   observes that  \begin{equation}\label{nonorthogonalC}
	\langle\psi_i|\psi_j\rangle_B\neq 0,\  \langle\psi_{k-1+i}|\psi_{k-1+j}\rangle_A\neq 0
\end{equation} 
for different $i,j\in\{0,1,\cdots,k-1\}.$
If the set is local redundancy,  there must exist not all empty sets $\mathcal{S}_A\subseteq \{A_1,A_2,\cdots,A_n\}$, $\mathcal{S}_B\subseteq \{B_1,B_2,\cdots,B_n\}$   and  some  unitaries $U_A,U_B\in U(d),$ 
such that 
$$\{\mathrm{Tr}_{\mathcal{S}_A \cup \mathcal{S}_B }[(U_A\otimes U_B)|\psi_i\rangle\langle \psi_i|(U_A^\dagger\otimes U_B^\dagger)]\}_{i=1}^{2k-2}$$  are mutually orthogonal. And this  is equivalent to the mutual  orthogonality of the following set
$$\{\mathrm{Tr}_{\mathcal{S}_A  }[U_A |\psi_i\rangle_A\langle \psi_i|U_A^\dagger ] \otimes \mathrm{Tr}_{\mathcal{S}_B  }[U_B |\psi_i\rangle_B\langle \psi_i|U_B^\dagger ]\}_{i=1}^{2k-2}.$$
To ensure  the satisfaction of these orthogonality relations, by Eq. \eqref{nonorthogonalC} and  Lemma  \ref{lemma:nonorthogonal},  both  $\{\mathrm{Tr}_{\mathcal{S}_A  }[U_A |\psi_i\rangle_A\langle \psi_i|U_A^\dagger ]\}_{i=1}^{k-1}$  and $\{\mathrm{Tr}_{\mathcal{S}_B  }[U_B |\psi_{k-1+i}\rangle_B\langle \psi_{k-1+i}|U_B^\dagger ]\}_{i=1}^{k-1}$    are mutually orthogonal. By assumption, at least one of $\mathcal{S}_A$ and  $\mathcal{S}_B$ is nonempty. Without loss of generality, we assume that $\mathcal{S}_A\neq \emptyset.$ Therefore, 
the resulting states $\{\mathrm{Tr}_{\mathcal{S}_A  }[U_A |\psi_i\rangle_A\langle \psi_i|U_A^\dagger ]\}_{i=1}^{k-1}$ are $(k-1)$ orthogonal states in the systems corresponding to $\{A_1,\cdots,A_n\}\setminus \mathcal{S}_A$ whose dimension is at most $d/p_1\leq d/3<k-1$. This deduces a contradiction as there are at most $N$ orthogonal positive semidefinite matrices in system with dimensional $N$. Therefore, our constructing set is locally irredundant.

	\begin{thm}
	Let $d\geq 11$ be an odd  integer.  Then there exists some orthogonal set without entanglement  in $\mathbb{C}^d\otimes\mathbb{C}^d$ which has  genuine hidden nonlocality of type I .   
\end{thm}
	
	One finds that the above results can be also use to construct sets with this property in multipartite systems. For example, let $\mathcal{H}=\mathcal{H}_A\otimes \mathcal{H}_B\otimes \mathcal{H}_C\otimes\mathcal{H}_D\otimes \mathcal{H}_E\otimes\mathcal{H}_F$ where   $$\dim_{\mathbb{C}}(\mathcal{H}_A ) =\dim_{\mathbb{C}}(\mathcal{H}_B ) =\dim_{\mathbb{C}}(\mathcal{H}_C ) =\dim_{\mathbb{C}}(\mathcal{H}_D )=11$$
  and $\dim_{\mathbb{C}}(\mathcal{H}_E ) =\dim_{\mathbb{C}}(\mathcal{H}_F )=13.$  Set $\mathbb{M}^B:=\{\pi_1^B= \sum_{i=0}^4|i\rangle\langle i|,\ \pi_2^B=  \sum_{j=5}^{10}|j\rangle\langle j|\}.$ Denote $\{|\Phi_i\rangle_{AB}\}_{i=1}^{20}$ be the set constructed in Example  \ref{ex-gan11}.  Finding four unfilled  squares (with labels $h_i,H_i,v_j,V_j$) in Fig. \ref{fig:example1_11} such that they come from exactly four different colored blocks and form a rectangle. For example, $\{(3,2),(3,8),(9,2),(9,8)\}.$  Using the four cubics, we can define a state  $|\Phi\rangle_{AB}:=|3-9\rangle|2-8\rangle$ with the following properties: (1). The set $\{|\Phi_i\rangle_{AB}\}_{i=1}^{20} \cup \{|\Phi\rangle_{AB}\}$ is  an orthogonal set  which is still locally distinguishable (whose proof is similar to the proof showed in Example \ref{ex-gan11});  (2). $\pi_k^B|\Phi\rangle_{AB} $  is nonzero for each $k=1,2$; (3).   $\pi_k^B|\Phi\rangle_{AB}$  is orthogonal to each of 
	$\{\pi_k^B|\Phi_i\rangle_{AB}\}_{i=1}^{20}$for  $k=1,2$.      Denote $\{|\Psi_j\rangle_{EF}\}_{j=1}^{24}$ be the set constructed in Example  \ref{ex-gan13}. And we choose $|\Psi\rangle_{EF}$  to be $|4-10\rangle|3-9\rangle$  which has similar properties with $|\Phi\rangle_{AB}$.   One can check that the union of the following set 
$$	\begin{array}{l}
		\{|\Phi_i\rangle_{AB}|\Phi\rangle_{CD}|\Psi\rangle_{EF}\}_{i=1}^{20},\\[2mm] \{|\Phi\rangle_{AB}|\Phi_j\rangle_{CD}|\Psi\rangle_{EF}\}_{j=1}^{20},\\[2mm] \{|\Phi\rangle_{AB}|\Phi\rangle_{CD}|\Psi_k\rangle_{EF}\}_{k=1}^{24}, 
		\end{array}$$
	has genuine hidden nonlocality of type I   in $\mathcal{H}=\mathcal{H}_A\otimes \mathcal{H}_B\otimes \mathcal{H}_C\otimes\mathcal{H}_D\otimes \mathcal{H}_E\otimes\mathcal{H}_F$. Firstly, the set is locally distinguishable as  $\{|\Phi_i\rangle_{AB}\}_{i=1}^{20} \cup \{|\Phi\rangle_{AB}\}$, $\{|\Phi_i\rangle_{CD}\}_{j=1}^{20} \cup \{|\Phi\rangle_{CD}\}$  and $\{|\Psi_i\rangle_{EF}\}_{k=1}^{24} \cup \{|\Psi\rangle_{EF}\}$  are all locally distinguishable.    Now if the $B,D,F$ parts make the following local measurements 
$$	\begin{array}{l}
	\mathbb{M}^B:=\{\pi_1^B= \sum_{i=0}^4|i\rangle\langle i|,\ \pi_2^B=  \sum_{j=5}^{10}|j\rangle\langle j|\},\\[1mm]
		\mathbb{M}^D:=\{\pi_1^D= \sum_{i=0}^4|i\rangle\langle i|,\ \pi_2^D=  \sum_{j=5}^{10}|j\rangle\langle j|\},\\[1mm]
			\mathbb{M}^F:=\{\pi_1^F= \sum_{i=0}^5|i\rangle\langle i|,\ \pi_2^F=  \sum_{j=6}^{12}|j\rangle\langle j|\}
\end{array}$$ 
there are eight outcomes. One can check that the post-measurement states of each outcome  form  an orthogonal set with $64$ elements which is locally indistinguishable. Generally, this result can be extended to $\bigotimes_{i=1}^N (\mathcal{H}_{A_i}\otimes \mathcal{H}_{B_i} )$ where each  $d_i=\mathrm{dim}_{\mathbb{C}}(\mathcal{H}_{A_i})=\mathrm{dim}_{\mathbb{C}}(\mathcal{H}_{B_i})$ is an odd integer greater than 10 for  $i\in \{1,2,\cdots,N\}$.

	\vskip 5pt 
	
		Note that locally irreducible (see \cite{Halder19}) is a stronger form of nonlocality than locally indistinguishable. Therefore, it is interesting to find that whether can we genuinely activate this stronger form of nonlocality under OPLM without decreasing the cardinality of the given set.
		The examples with three elements provided by \cite{Bandyopadhyay21} do satisfy this property. In the following, we present an example of such set without entanglement.
	
	\begin{figure}[h]
	\centering
	\includegraphics[width=0.48\textwidth,height=0.48\textwidth]{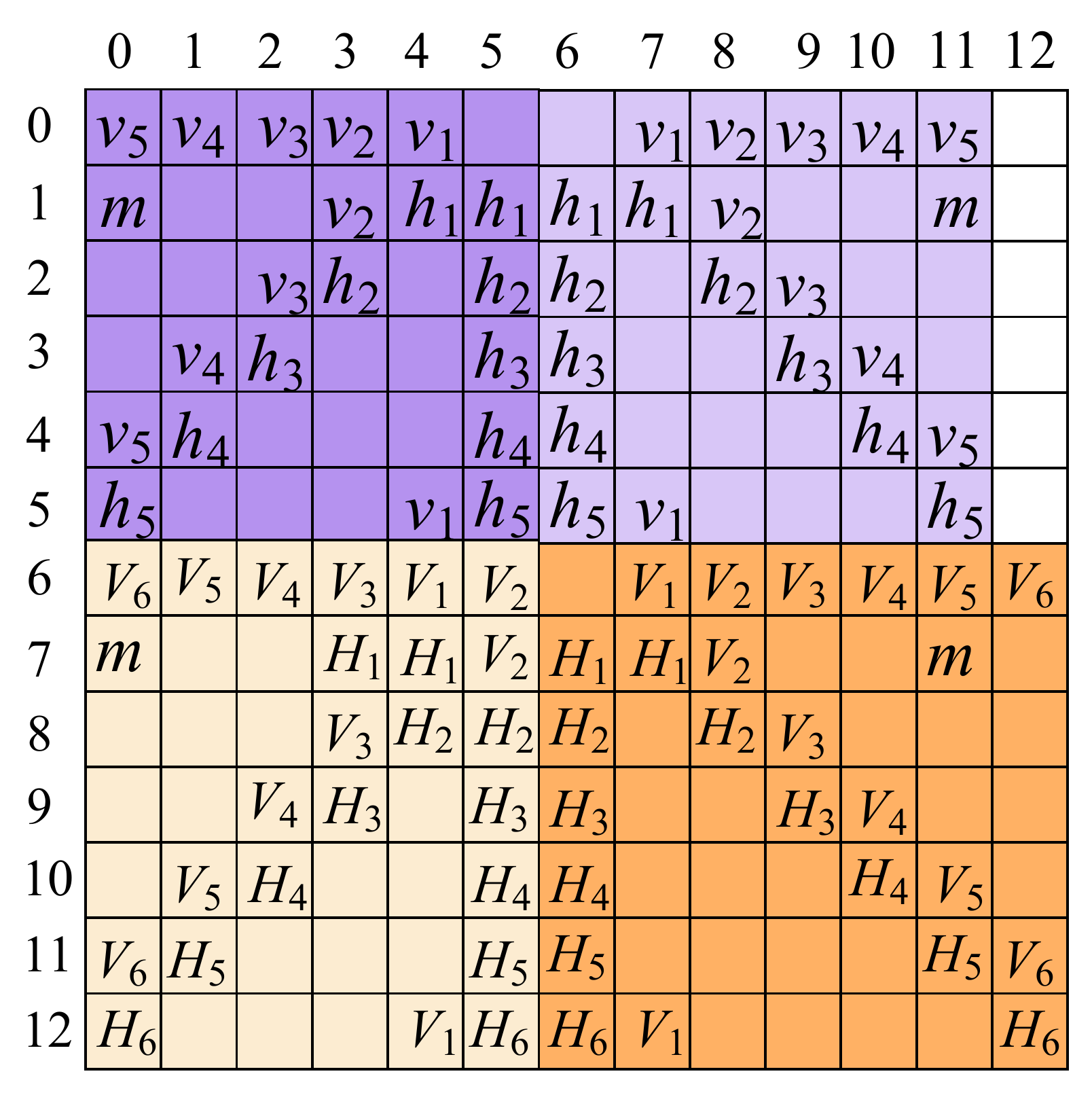}
\caption{This shows the  states structure of set of product states in Example  \ref{ex-Irr11} .  The notation here is similar with that in Fig. \ref{fig:example1_11}.  The four squares ($(1,0), (1,9), (6,0),(6,9)$  labeled with `$m$' correspond to the states $|\mathcal{M}\rangle=|1-6\rangle|0-9\rangle$.}\label{fig:example1_Irr11}
\end{figure}
	
	\begin{example}
		\label{ex-Irr11} 
		Let $\mathcal{S}$ be the set of product states in $\mathbb{C}^{11}\otimes\mathbb{C}^{11}$ whose elements are listed as below (see Fig. \ref{fig:example1_Irr11})):
		\begin{equation*}	
			\begin{array}{rclrcl}
				\left|\psi_{1}\right\rangle  & \equiv & |1\rangle|3- 4 +5-6\rangle,& 	\left|\psi_{5}\right\rangle  & \equiv & |0-4\rangle|3-6\rangle,\\ \left|\psi_{2}\right\rangle  & \equiv & |2\rangle|2-4 +5-7\rangle, & \left|\psi_{6}\right\rangle  & \equiv &| 0-1\rangle|2-7\rangle,\\
				\left|\psi_{3}\right\rangle  & \equiv & |3\rangle|1-4+5-8\rangle,&  \left|\psi_{7}\right\rangle  & \equiv & |0-2\rangle|1-8\rangle,\\	\left|\psi_{4}\right\rangle  & \equiv &| 4\rangle|0-4+5-9\rangle,
				& 	\left|\psi_{8}\right\rangle  & \equiv &| 0-3\rangle|0-9\rangle,\\
				\left|\phi_{1}\right\rangle  & \equiv & |6\rangle|5-6+2-3\rangle,& 	\left|\phi_{6}\right\rangle  & \equiv & |5-10\rangle|6-3\rangle,\\ \left|\phi_{2}\right\rangle  & \equiv & |7\rangle|5-7+3-4\rangle, & \left|\phi_{7}\right\rangle  & \equiv &| 5-6\rangle|7-4\rangle,\\
				\left|\phi_{3}\right\rangle  & \equiv & |8\rangle|5-8+2-4\rangle,&  \left|\phi_{8}\right\rangle  & \equiv & |5-7\rangle|8-2\rangle,\\	\left|\phi_{4}\right\rangle  & \equiv &| 9\rangle|5-9+1-4\rangle,
				& 	\left|\phi_{9}\right\rangle  & \equiv &| 5-8\rangle|9-1\rangle,	\\
				\left|\phi_{5}\right\rangle  & \equiv &| 10\rangle|5-10+0-4\rangle,
				& 	\left|\phi_{10}\right\rangle  & \equiv &| 5-9\rangle|10-0\rangle,	
				
			\end{array}
		\end{equation*}
		and  $|S\rangle\equiv |+_{10}\rangle|+_{10}\rangle,$   and $|\mathcal{M}\rangle\equiv |1-6\rangle|0-9\rangle.$		  
		Then the set $\mathcal{S}$ has genuine hidden strong form nonlocality of type I. 
	\end{example} 
Similar with Example \ref{ex-gan11}, we can show that $\mathcal{S}$ is locally distinguishable.
  Suppose Bob perform the measurement
	$\mathbb{M}^B:=\{\pi_1^B= \sum_{i=0}^4|i\rangle\langle i|,\ \pi_2^B=  \sum_{j=5}^{10}|j\rangle\langle j|\}.$ Note that, for both outcomes of the measurement, the cardinaility of the possible states are unchanged. In appendix, we show that for both outcomes of the measurement the postmeasurement states  are locally irreducible.

	\section{Genuine hidden nonlocality of type II}\label{sec:Decreasing}

	In this section, we study how to construct sets without entanglement but have genuine hidden nonlocality of type II. We will use the local  indistinguishability of  the  sets  without entanglement as follows.

	\begin{thm}[\cite{Xu20a}]\label{thm:CP_Nonlocality}
		In $\mathbb{C}^m\otimes \mathbb{C}^n$ system with $m,n\geq 3$, there exists  a set with $2(m+n)-4$   orthogonal pure product states  which is  locally indistinguishable.
	\end{thm}
	\begin{figure}[h]
		\centering
		\includegraphics[width=0.49\textwidth,height=0.43\textwidth]{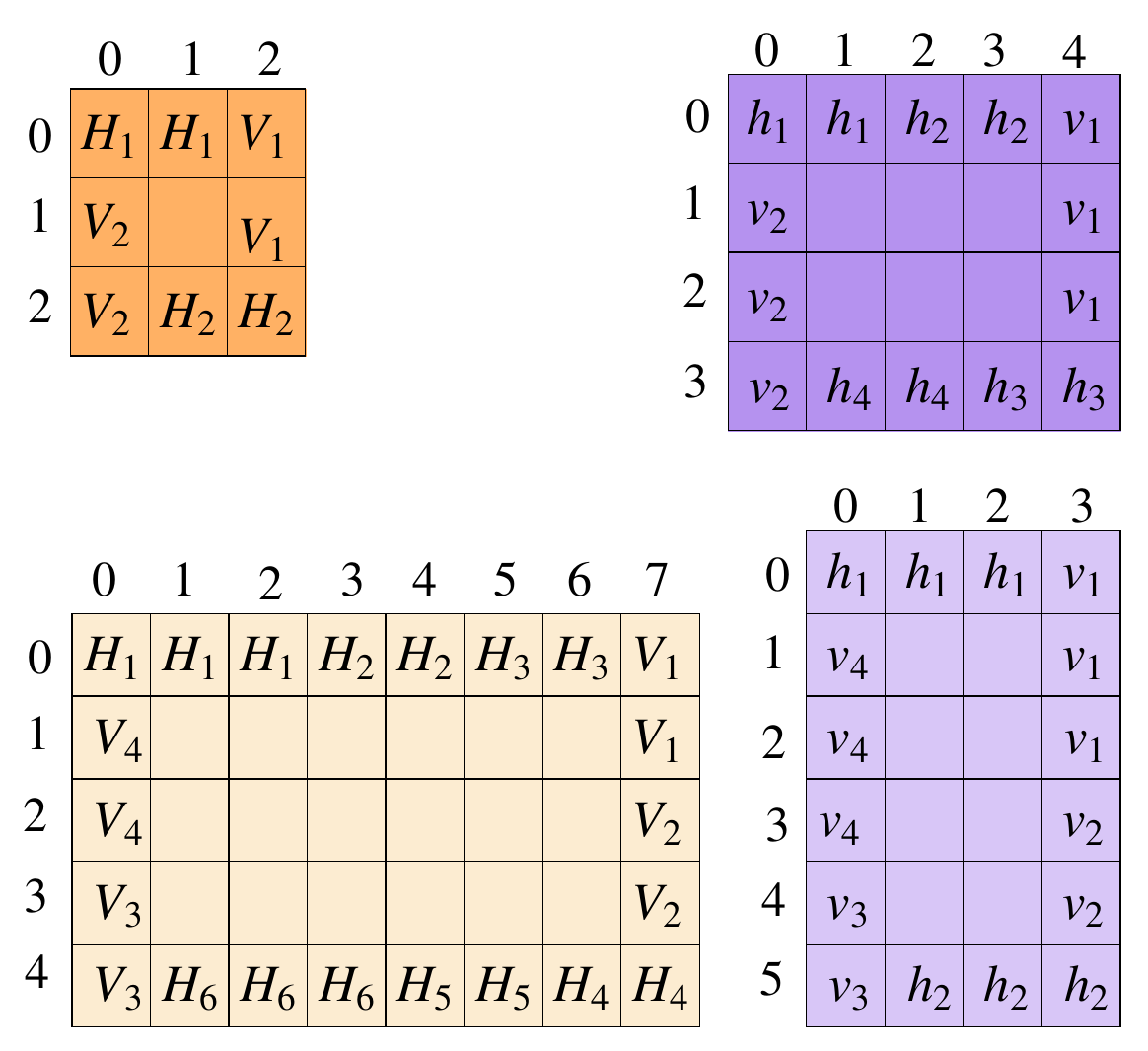}
		\caption{This shows the examples of  the four types of states structure in Theorem \ref{thm:CP_Nonlocality} in  $\mathbb{C}^3\otimes \mathbb{C}^3$, $\mathbb{C}^4\otimes \mathbb{C}^5$, $\mathbb{C}^5\otimes \mathbb{C}^8$, and $\mathbb{C}^6\otimes \mathbb{C}^4$  respectively. The squares  indicated by the same label  of cardinality $2,3$ represent  $2,3$ states respectively.  For example, there are three squares (that is, $(0,3), (1,3),$ and $(2,3)$) with label `$v_1$' at the right down corner figure, they correspond to  $|0+w1+w^2 2\rangle|3\rangle\in \mathbb{C}^6\otimes \mathbb{C}^4$ where  $w\in\{1,e^{\frac{2\pi \sqrt{-1}}{3}},e^{\frac{4\pi \sqrt{-1}}{3}}\}$). }\label{fig:Fourtypes}
	\end{figure}
	
	According to the parity of $m$ and $n $, the structure of the set of states can be divided into four types (see Fig. \ref{fig:Fourtypes}).
	For more details on the sets of quantum states, the authors may see  the reference  \cite{Xu20a}. 	
	Now let's start with  the following example.

	\begin{example}
		\label{ex-7-8} 
		Let $\mathcal{S}$ be the set of product states in $\mathbb{C}^{7}\otimes\mathbb{C}^{8}$ whose elements are listed as below:
		{\small	$$\begin{array}{rclrcl}
				\left|\psi_{1,2}\right\rangle&\equiv& |0\rangle|0\pm 1\rangle,&  \left|\psi_{8,9}\right\rangle& \equiv&  |3\rangle|3\pm 4+5\pm 6\rangle,\\	 
				\left|\psi_{3,4}\right\rangle& \equiv&  |0\rangle|2\pm 3\rangle,
				&  			\left|\psi_{10,11}\right\rangle& \equiv&  |3\rangle|1\pm 2\rangle, \\ 
				\left|\psi_{5,6,7}\right\rangle& \equiv&  |0+w1+w^22\rangle|4\rangle,
				& \left|\psi_{12,13,14}\right\rangle& \equiv&  |1+w2+w^23\rangle|0\rangle, \\			
				\left|\phi_{1,2}\right\rangle  &\equiv & |4\rangle|5\pm6+3\pm 4\rangle, & \left|\phi_{5,6}\right\rangle  &\equiv & |6\rangle|6\pm7 \rangle, 
				\    \\
				\left|\phi_{3,4}\right\rangle & \equiv& |  4\pm 5\rangle|7\rangle,
				&
				\   \left|\phi_{7,8}\right\rangle & \equiv& | 5\pm 6\rangle|5\rangle,    
			\end{array}$$}
		where $w\in\{1,e^{\frac{2\pi \sqrt{-1}}{3}},e^{\frac{4\pi \sqrt{-1}}{3}}\}$ which is any cubic root of unit.	 	Then the set $\mathcal{S}$ has genuine hidden nonlocality of type II. 
	\end{example} 
	
	\begin{figure}[h]
		\centering
		\includegraphics[width=0.48\textwidth,height=0.32\textwidth]{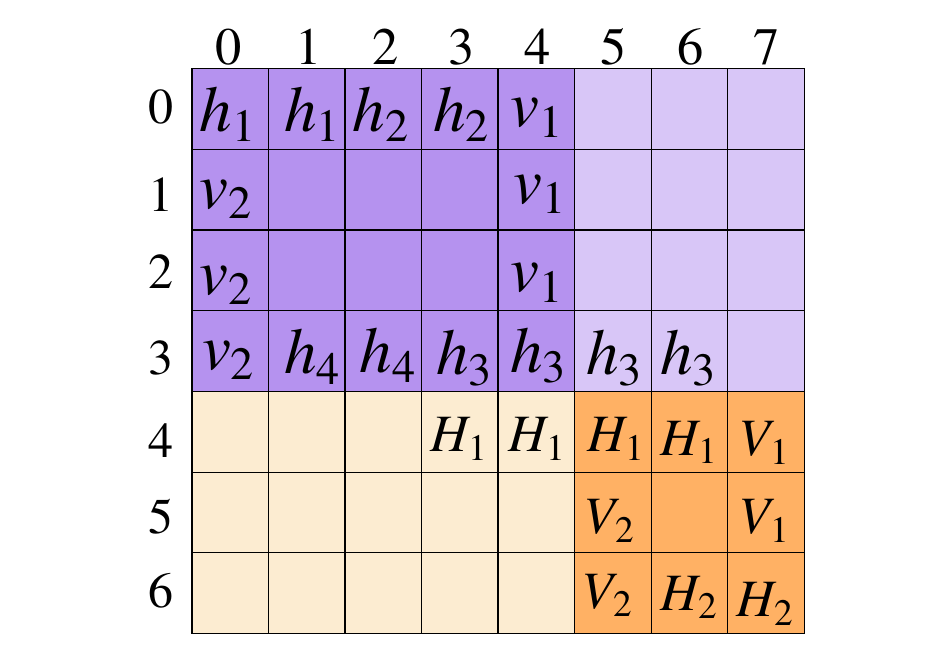}
		\caption{This shows the  states structure of set of product states in $\mathbb{C}^{7}\otimes\mathbb{C}^{8}$ of Example \ref{ex-7-8}. The squares  indicated by the same label  of cardinality $2,3,4$ represent  $2,3,2$ states respectively. For examples, there are two squares (that is, $(0,2), (0,3)$) with label `$h_2$', they correspond to two states $|\psi_{3,4}\rangle:=|0\rangle|2\pm 3\rangle,$    and  there are four squares (that is, $  (3,3),(3,4),$ $(3,5)$, and $(3,6)$) with label `$h_3$', they correspond to two states $|\phi_{8,9}\rangle:=|3\rangle|3\pm4+5\pm 6\rangle $  (More exactly, $|\phi_{8}\rangle:=|3\rangle|3+4+5+6\rangle$ and $|\phi_{9}\rangle:=|3\rangle|3-4+5-6\rangle$).}\label{fig:states7-8}
	\end{figure}
	
	Now we show that $\mathcal{S}$ is locally distinguishable. First, Alice perform the measurement $\mathbb{M}_1^A:=\{\pi_1^A=\sum_{i=0}^3|i\rangle\langle i |, \ \pi_2^A= \sum_{j=4}^6|j\rangle\langle j |\}.$
	
	If the outcome  of $\mathbb{M}_1^A$ is `1', the state must be one of $\{|\psi_i\rangle\}_{i=1}^{14}.$ And the states do not change under this measurement. Now Bob perform the measurement:
	$$\mathbb{M}_1^B:=\{\pi_1^B= |4\rangle\langle 4|+|5\rangle\langle 5|,\ \pi_2^B= \mathbb{I}_8-\pi_i^B\}.$$
	If the outcome  of $\mathbb{M}_1^B$ is `1', the state must be one of $\{|\psi_{5,6,7,8,9}\rangle\}$ which  may change to be 
	\begin{eqnarray*}
		\left|\psi_{5,6,7}\right\rangle=  |0+w1+w^22\rangle|4\rangle,\
		\left|\tilde{\psi}_{8,9}\right\rangle = |3\rangle|5\pm 4\rangle. 
	\end{eqnarray*}
	It is easy to derive a LOCC protocal to distinguish the states of above set. If the outcome of $\mathbb{M}_1^B$ is `2', the state must be one of $\{|\psi_{i}\rangle\}_{i=1}^{14}\setminus \{|\psi_{5,6,7}\rangle\}$ (without the states corresponding to the label $v_1$ in Fig. \ref{fig:states7-8}) which  may change to   
	{\small	$$\begin{array}{rclrcl}
			\left|\psi_{1,2}\right\rangle&=& |0\rangle|0\pm 1\rangle,&  \left|\tilde{\psi}_{8,9}\right\rangle& =&  |3\rangle|3\pm 6\rangle,\\	 
			\left|\psi_{3,4}\right\rangle& =&  |0\rangle|2\pm 3\rangle,
			&  			\left|\psi_{10,11}\right\rangle& =&  |3\rangle|1\pm 2\rangle, \\ 
			&  &  
			& \left|\psi_{12,13,14}\right\rangle& =&  |1+w2+w^23\rangle|0\rangle.
		\end{array}$$	
		It is also easy to derive a LOCC protocal to distinguish the states of above set.
		
		If the outcome of $\mathbb{M}_1^A$   is `2', the state must be one of $\{|\phi_j\rangle\}_{j=1}^{8}.$ And the states do not change under this measurement. 
		Now Bob perform the measurement:
		$$\mathbb{M}_1^B:=\{\pi_1^B= |4\rangle\langle 4|+|5\rangle\langle 5|,\ \pi_2^B= \mathbb{I}_8-\pi_i^B\}.$$
		If the outcome of $\mathbb{M}_1^B$ is `1', the state must be one of $\{|\phi_{1,2,7,8}\rangle\}$ which  may change to be 
		\begin{eqnarray*}
			\left|\tilde{\phi}_{1,2}\right\rangle  = |4\rangle|5\pm 4\rangle ,\ 
			\left|\phi_{7,8}\right\rangle =|5\pm 6\rangle|5\rangle.
		\end{eqnarray*}
		It is easy to derive a LOCC protocal to distinguish these four states. If the outcome  of $\mathbb{M}_1^B$ is `2', the state must be one of $\{|\phi_{i}\rangle\}_{i=1}^8\setminus\{|\phi_{7,8}\rangle\}$ (without the states corresponding to the label $V_2$ in Fig. \ref{fig:states7-8}) which  may change to be $$\begin{array}{rclrcl} \left|\phi_{1,2}\right\rangle  & = & |4\rangle|3\pm6\rangle, & \left|\phi_{5,6}\right\rangle  &= & |6\rangle|6\pm7 \rangle,
			\    \\
			\left|\phi_{3,4}\right\rangle & =& |  4\pm 5\rangle|7\rangle.
			&
			&  &    
		\end{array}$$
		It is also easy to derive a LOCC protocal to distinguish the states of above set.

		Now we prove the second property. Suppose Bob perform the measurement
		$\mathbb{M}^B:=\{\pi_1^B= \sum_{i=0}^4|i\rangle\langle i|,\ \pi_2^B=  \sum_{j=5}^7|j\rangle\langle j|\}.$
		
		If the outcome of $\mathbb{M}^B$ is `1', the states are transferred 
		to 	
		{\small	$$\begin{array}{rclrcl}
				\left|\psi_{1,2}\right\rangle&=& |0\rangle|0\pm 1\rangle,&  \left|\tilde{\psi}_{8,9}\right\rangle& =&  |3\rangle|3\pm 4 \rangle,\\	 
				\left|\psi_{3,4}\right\rangle& =&  |0\rangle|2\pm 3\rangle,
				&  			\left|\psi_{10,11}\right\rangle& =&  |3\rangle|1\pm 2\rangle, \\ 
				\left|\psi_{5,6,7}\right\rangle& =&  |0+w1+w^22\rangle|4\rangle,
				& \left|\tilde{\psi}_{12,13,14}\right\rangle& =&  |1+w2+w^23\rangle|0\rangle, \\			
				\left|\phi_{1,2}\right\rangle  &= & |4\rangle| 3\pm 4\rangle, &    & & 		    
			\end{array}$$}
		which are mutually orthogonal and contains $14$ states that  is known to be  locally indistinguishable in $\mathbb{C}^4\otimes  \mathbb{C}^5$. 
		
		If the outcome  of $\mathbb{M}^B$ is `2', the states are transferred 
		to 	
		{ 	$$\begin{array}{rclrcl}
				& &  &  \left|\tilde{\psi}_{8,9}\right\rangle& \equiv&  |3\rangle|5\pm 6\rangle,\\	 		
				\left|\tilde{\phi}_{1,2}\right\rangle  &\equiv & |4\rangle|5\pm6 \rangle, & \left|\phi_{5,6}\right\rangle  &\equiv & |6\rangle|6\pm7 \rangle, 
				\    \\
				\left|\phi_{3,4}\right\rangle & \equiv& |  4\pm 5\rangle|7\rangle,
				&
				\   \left|\phi_{7,8}\right\rangle & \equiv& | 5\pm 6\rangle|5\rangle,    
			\end{array}$$}
		which are mutually orthogonal and contains $8$ states that  is known to be  locally indistinguishable in $\mathbb{C}^3\otimes  \mathbb{C}^3$.

		Now we prove that the set is locally irredundant. Written $|\psi_i\rangle=|\psi_i\rangle_A|\psi_i\rangle_B$ for $i=1,2,\cdots, 14.$
		One may observe that 
		$$\langle \psi_i|\psi_j\rangle_A\neq 0, \text{ and }  \langle \psi_{k}|\psi_{l}\rangle_B\neq 0$$
		for different $i,j\in\{1,2,3,4,5\}$ and  different $k,l\in\{1,12,13,14\}.$  If the set $\mathcal{S}$ is local redundancy,  there must exist  $\mathcal{S}_B\subseteq \{B_1,B_2,B_3\}$   and  some unitary $U_B\in U(d),$ 
		such that 
		$ \{\mathrm{Tr}_{\mathcal{S}_B  }[U_B |\psi_i\rangle_B\langle \psi_i|U_B^\dagger ]\}_{i=1}^{5}$ are mutually orthogonal. However, the dimension of the  system corresponding to $\{B_1,B_2,B_3\}\setminus \mathcal{S}_B$ is at most $4$ which is less than $5$.		\qed
		
		\vskip 5pt 
		
		The above argument can be easily extend to $\mathbb{C}^{m}\otimes \mathbb{C}^{n}$ provided the conditions $m,n\geq 7$. Therefore, we arrive at the following theorem.
		
		\begin{thm}
			Let $m,n\geq 7$ be an   integer. Then there exists some orthogonal set without entanglement in $\mathbb{C}^m\otimes\mathbb{C}^n$ which  has genuine hidden nonlocality of type II .   
		\end{thm}
		For an integer $m\geq 7$ define $f_m=\lfloor \frac{m-1}{2}\rfloor$ and $c_m=m-f_m$. Similarly, one finds that $c_m$ is the smallest integer which is larger than $m/2$.  Similarly, one can construct a set with $2(m+n)-8$ states which  are inspired by Xu et al.'s results \cite{Xu20a} in  $\mathbb{C}^{c_m}\otimes \mathbb{C}^{c_n}$ and   $\mathbb{C}^{f_m}\otimes \mathbb{C}^{f_n}.$ The local irredundancy can be obtained similarly as the above argument since there are at least $c_n>\frac{n}{2}$ states for which Alice's part are mutually non-orthogonal and   $c_m>\frac{m}{2}$ states for which Bob's part are mutually non-orthogonal. Similar with the discussion in the end of  Sec. \ref{sec:Stable}, we can extend the construction to multipartite systems whose local dimensions are greater than $6$.

		\section{Conclusions and Discussions}\label{sec:Con}	
		We studied genuine hidden nonlocality without entanglement  in the settings of local discrimination of quantum states. There are two cases: the first one is cardinality preserving under OPLM and the second one is cardinality decreasing under OPLM.  For both cases, we provided   examples  to demonstrate the idea   how to construct  set of product states with genuine hidden nonlocality. For the first case, the construction can even be extended to  $2N$-parties systems $\bigotimes_{i=1}^N (\mathcal{H}_{A_i}\otimes \mathcal{H}_{B_i} )$ whenever each  $d_i=\mathrm{dim}_{\mathbb{C}}(\mathcal{H}_{A_i})=\mathrm{dim}_{\mathbb{C}}(\mathcal{H}_{B_i})$ is an odd integer greater than 10 for  $i\in \{1,2,\cdots,N\}.$ And for the second case, we can extend the idea to any $N$-parties systems whose local dimensions are greater than $6$. Moreover, we provided a method to deal with the local redundancy problem which might be helpful to tackling this problem.  Compared with results with those in \cite{Bandyopadhyay21}, it is a bit surprise as  sets without entanglement seems can be more easier to be genuinely activated than  sets with  entanglement. Therefore, it is   interesting  find more sets with entanglement (but with more complicated structure) and genuine hidden nonlocality. In this case, finding a way to check   the local irredundancy is a tricky problem.

		\begin{acknowledgments}
		We thank Shi Fei for pointing out several mistakes of our orginal maniscript.	This  work  is supported  by  National  Natural  Science  Foundation  of  China under Grant No. 12005092,  the China Postdoctoral Science Foundation (2020M681996), the Key Research and Development Project of Guangdong province under Grant
			No. 2020B0303300001, the Guangdong Basic and Applied
			Research Foundation under Grant No. 2020B1515310016. 
		\end{acknowledgments}
		
		\vskip 10pt

		{
			$${\text{\textbf{APPENDIX  }}}$$
		}
		
		\noindent{\bf Proof of Lemma \ref{lemma:nonorthogonal}.}  Suppose not, we have $\langle \mathcal{N}(\rho), \mathcal{N}(\sigma)\rangle_B=0.$ 
		Let $\rho=\sum_{i=1}^m\lambda_i |x_i\rangle\langle x_i|, \sigma=\sum_{j=1}^n\mu_j |y_j\rangle\langle y_j|$ be the spectral decomposition of  $\rho$ and $\sigma$ respectively.
		By the linearity of $\mathcal{N}$ and the inner product,
		\begin{equation}\label{eq:zero}
			\langle \mathcal{N}(\rho), \mathcal{N}(\sigma)\rangle_B=\sum_{i=1}^m\sum_{j=1}^n \lambda_i \mu_j \langle \mathcal{N}(|x_i\rangle\langle x_i|), \mathcal{N}(|y_j\rangle\langle y_j|)\rangle_B.
		\end{equation}
		It is well known that $\mathrm{Tr}[M N]\geq 0$ if $M$ and $N$ are positive semidefinite  matrices. Therefore, the zero of $\langle \mathcal{N}(\rho), \mathcal{N}(\sigma)\rangle_B$ implies all the terms on the right hand side of  Eq. \eqref{eq:zero} are zeros, i.e., 
		$$\langle \mathcal{N}(|x_i\rangle\langle x_i|), \mathcal{N}(|y_j\rangle\langle y_j|)\rangle_B=0$$
		for all $i,j$. 
		For each $i,j$, we compute the $ \langle \mathcal{N}(|x_i\rangle\langle x_i|), \mathcal{N}(|y_j\rangle\langle y_j|)\rangle_B $ by using the Kraus representation of $\mathcal{N}$  which is given by 
		{\small $$\mathcal{N}(\tau)=\sum_{k=1}^K A_k \tau A_k^\dagger, \ \text{ where }  \sum_{k=1}^KA_k^\dagger A_k=\mathbb{I}_A.$$}
		Therefore, we have 
		\begin{equation*}
			\begin{array}{rcl}
				\mathcal{N}(|x_i\rangle\langle x_i|), \mathcal{N}(|y_j\rangle\langle y_j|)\rangle_B&=&\displaystyle \sum_{k=1}^K\sum_{l=1}^K  \mathrm{Tr}[A_k |x_i\rangle\langle x_i| A_k^\dagger A_l |y_j\rangle\langle y_j| A_l^\dagger]\\[2mm]
				&=&\displaystyle \sum_{k=1}^K\sum_{l=1}^K  \mathrm{Tr}[\langle x_i| A_k^\dagger A_l |y_j\rangle\langle y_j| A_l^\dagger A_k |x_i\rangle]\\[2mm]
				&=&\displaystyle \sum_{k=1}^K\sum_{l=1}^K  |\langle x_i| A_k^\dagger A_l |y_j\rangle|^2.			
			\end{array}
		\end{equation*}
		As $\langle \mathcal{N}(|x_i\rangle\langle x_i|), \mathcal{N}(|y_j\rangle\langle y_j|)\rangle_B=0$, we have $|\langle x_i| A_k^\dagger A_l |y_j\rangle|^2=0$ for all $1\leq k,l\leq K$. Particularly,  $\langle x_i| A_k^\dagger A_k |y_j\rangle=0$ for $1\leq k\leq K$ which imply that $\langle x_i|y_j\rangle= 0$ as  $\sum_{k=1}^KA_k^\dagger A_k=\mathbb{I}_A.$ Using these results and the spectral decomposition of $\rho$ and $\sigma$, we obtain that $\rho\sigma=\mathbf{0}$ which is a zero matrix. This is contradicted with the non-orthogonality of $\rho$ and $\sigma$. \qed

		\vskip 5pt
		
	\noindent	{\bf Proof of the local irreducibility of postmeasurement states in Example \ref{ex-Irr11}.} 	If the outcome of $\mathbb{M}^B$ is `1', the states are transferred 
		to 	 	$\tilde{\mathcal{S}}_1$
		\begin{equation*}	
			\begin{array}{rclrcl}
				\left|\tilde{\psi}_{1}\right\rangle  & = & |1\rangle|3- 4 \rangle,& 	\left|\tilde{\psi}_{5}\right\rangle  & = & |0-4\rangle|3\rangle,\\ \left|\tilde{\psi}_{2}\right\rangle  & = & |2\rangle|2-4 \rangle, & \left|\tilde{\psi}_{6}\right\rangle  & = &| 0-1\rangle|2\rangle,\\
				\left|\tilde{\psi}_{3}\right\rangle  & = & |3\rangle|1-4\rangle,&  \left|\tilde{\psi}_{7}\right\rangle  & = & |0-2\rangle|1\rangle,\\	\left|\tilde{\psi}_{4}\right\rangle  & =&| 4\rangle|0-4\rangle,
				& 	\left|\tilde{\psi}_{8}\right\rangle  & = &| 0-3\rangle|0\rangle,\\
				\left|\tilde{\phi}_{1}\right\rangle  & = & |6\rangle|2-3\rangle,& 	\left|\tilde{\phi}_{6}\right\rangle  & = & |5-10\rangle|3\rangle,\\ \left|\tilde{\phi}_{2}\right\rangle  & = & |7\rangle|3-4\rangle, & \left|\tilde{\phi}_{7}\right\rangle  & = &| 5-6\rangle|4\rangle,\\
				\left|\tilde{\phi}_{3}\right\rangle  & = & |8\rangle|2-4\rangle,&  \left|\tilde{\phi}_{8}\right\rangle  & = & |5-7\rangle|2\rangle,\\	\left|\tilde{\phi}_{4}\right\rangle  & = &| 9\rangle|1-4\rangle,
				& 	\left|\tilde{\phi}_{9}\right\rangle  & = &| 5-8\rangle|1\rangle,	\\
				\left|\tilde{\phi}_{5}\right\rangle  & = &| 10\rangle|0-4\rangle,
				& 	\left|\tilde{\phi}_{10}\right\rangle  & = &| 5-9\rangle|0\rangle,	\\				\left|\tilde{S}\right\rangle  & = &| +_{10}\rangle|+_4\rangle,
			& 	 
			\left|\tilde{\mathcal{M}} \right\rangle  & = &|1-6\rangle|0\rangle, 		\end{array}
		\end{equation*}
		which can be seen as states in $\mathcal{H}_A\otimes \mathcal{H}_{B_1}:=\mathbb{C}^{11}\otimes \mathbb{C}^5$ with computation basis $\{|i\rangle|j\rangle\ \big| \ i\in \mathbb{Z}_{11},j\in\mathbb{Z}_5\}.$
		We only need to show that $A$ and $B_1$ can only start with trivial OPLMs. In fact, to preserve the orthogonality of $\{|\tilde{\psi}_i\rangle\}_{i=1}^{8}\cup\{|\tilde{S}\rangle\},$ $B_1$ can only start with trivial measurement. Now it is  sufficient to prove that if $E=(a_{i,j})_{i,j\in\mathbb{Z}_{11}}$ is an $11\times 11$ Hermitian matrix and
		\begin{equation}\label{eq:OR1}
		    \langle \Theta_1| E\otimes \mathbb{I}_{B_1}|\Theta_2\rangle=0,\ \ |\Theta_1\rangle\neq |\Theta_2\rangle\in \tilde{\mathcal{S}}_1,
		\end{equation} 
		then $E\propto \mathbb{I}_A.$ As the $B_1$ part of the states $\{|\tilde{\phi}_{i}\rangle\}_{i=1}^4\cup \{|\tilde{\psi}_{j}\rangle\}_{j=1}^5$, applying Eq. \eqref{eq:OR} to this subset, we obtain 
		$$ a_{k,l}=0,\ \  \forall\  k,l\in \mathbb{Z}_{11}\setminus\{0,5\}, \text{ and } k\neq l.$$
		Substitute $|\Theta_1\rangle$ and $ |\Theta_2\rangle$ in Eq. \eqref{eq:OR1} by  one of $\{|\tilde{\psi}_{i}\rangle\}_{i=1}^4 \cup\{|\tilde{\phi}_{j}\rangle\}_{j=2}^5$ and  $|\tilde{\phi}_{7}\rangle$ respectively, we obtain 
	$ a_{x,5}=a_{x,6}=0 \text{ for } x\in \{1,2,3,4,7,8,9,10\}.$ Now substitute $|\Theta_1\rangle$ and $ |\Theta_2\rangle$ in Eq. \eqref{eq:OR1} by   $|\tilde{\phi}_{1}\rangle$ and $|\tilde{\phi}_{6}\rangle$, we obtain that $a_{6,5}=a_{6,10}=0.$ Now applying Eq, \eqref{eq:OR1} to the following pairs of states $(|\tilde{\psi}_{5}\rangle,|\tilde{\psi}_{1}\rangle ),$ $(|\tilde{\psi}_{6}\rangle,|\tilde{\psi}_{2}\rangle ),$ $(|\tilde{\psi}_{7}\rangle,|\tilde{\psi}_{3}\rangle ),$ $(|\tilde{\psi}_{8}\rangle,|\tilde{\psi}_{4}\rangle ),$ $(|\tilde{\psi}_{5}\rangle,|\tilde{\phi}_{6}\rangle ),$ $(|\tilde{\psi}_{5}\rangle,|\tilde{\phi}_{2}\rangle ),$ $(|\tilde{\psi}_{6}\rangle,|\tilde{\phi}_{3}\rangle ),$ $(|\tilde{\psi}_{7}\rangle,|\tilde{\phi}_{4}\rangle ),$
	$(|\tilde{\psi}_{8}\rangle,|\tilde{\phi}_{5}\rangle ),$ and $(|\tilde{\psi}_{5}\rangle,|\tilde{\phi}_{6}\rangle )$, we get 
	$a_{0,x}=0$ for $x\in\mathbb{Z}_{11}\setminus\{0\}.$ Therefore, $E$ is a diagonal matrix. At last,  applying Eq.\eqref{eq:OR1} to one of $\{|\tilde{\psi}_{i}\rangle\}_{i=5}^8\cup \{|\tilde{\phi}_{j}\rangle\}_{j=6}^{10}\cup \{|\tilde{\mathcal{M}}\rangle\}$ and $|\tilde{S}\rangle,$  we have  $E\propto \mathbb{I}_A$.

	If the outcome of $\mathbb{M}^B$ is `2', the states are transferred 
		to 	$\tilde{\mathcal{S}}_2$ 	
				\begin{equation*}	
			\begin{array}{rclrcl}
				\left|\tilde{\psi}_{1}\right\rangle  & = & |1\rangle|5-6\rangle,& 	\left|\tilde{\psi}_{5}\right\rangle  & = & |0-4\rangle|6\rangle,\\ \left|\tilde{\psi}_{2}\right\rangle  & = & |2\rangle|5-7\rangle, & \left|\tilde{\psi}_{6}\right\rangle  & = &| 0-1\rangle|7\rangle,\\
				\left|\tilde{\psi}_{3}\right\rangle  & = & |3\rangle|5-8\rangle,&  \left|\tilde{\psi}_{7}\right\rangle  & =& |0-2\rangle|8\rangle,\\	\left|\tilde{\psi}_{4}\right\rangle  & = &| 4\rangle|5-9\rangle,
				& 	\left|\tilde{\psi}_{8}\right\rangle  & = &| 0-3\rangle|9\rangle,\\
				\left|\tilde{\phi}_{1}\right\rangle  & = & |6\rangle|5-6\rangle,& 	\left|\tilde{\phi}_{6}\right\rangle  & = & |5-10\rangle|6\rangle,\\ \left|\tilde{\phi}_{2}\right\rangle  & = & |7\rangle|5-7\rangle, & \left|\tilde{\phi}_{7}\right\rangle  & = &| 5-6\rangle|7\rangle,\\
				\left|\tilde{\phi}_{3}\right\rangle  & = & |8\rangle|5-8\rangle,&  \left|\tilde{\phi}_{8}\right\rangle  & = & |5-7\rangle|8\rangle,\\	\left|\tilde{\phi}_{4}\right\rangle  & = &| 9\rangle|5-9\rangle,
				& 	\left|\tilde{\phi}_{9}\right\rangle  & = &| 5-8\rangle|9\rangle,	\\
				\left|\tilde{\phi}_{5}\right\rangle  & = &| 10\rangle|5-10\rangle,
				& 	\left|\tilde{\phi}_{10}\right\rangle  & = &| 5-9\rangle|10\rangle,\\		\left|\tilde{S}\right\rangle  & = &| +_{10}\rangle|{}_5+_{10}\rangle, &	\left|\tilde{\mathcal{M}} \right\rangle  & = &|1-6\rangle|9\rangle, 
			\end{array}
		\end{equation*}
			which can be seen as states in $\mathcal{H}_A\otimes \mathcal{H}_{B_2}:=\mathbb{C}^{11}\otimes \mathbb{C}^6$ with computation basis $\{|i\rangle|j\rangle\ \big| \ i\in \mathbb{Z}_{11},j\in\{5,6, \cdots,10\}\}.$ We only need to show that $A$ and $B_2$ can only start with trivial OPLMs. In fact, to preserve the orthogonality of $\{|\tilde{\phi}_i\rangle\}_{i=1}^{10}\cup\{|\tilde{S}\rangle\},$ $B_2$ can only start with trivial measurement. Now it is  sufficient to prove that if $E=(a_{i,j})_{i,j\in\mathbb{Z}_{11}}$ is an $11\times 11$ Hermitian matrix and
		\begin{equation}\label{eq:OR}
		    \langle \Theta_1| E\otimes \mathbb{I}_{B_2}|\Theta_2\rangle=0,\ \ |\Theta_1\rangle\neq |\Theta_2\rangle\in \tilde{\mathcal{S}}_2,
		\end{equation} 
		then $E\propto \mathbb{I}_A.$ 
		Written $E=\left(\begin{array}{cc}E_{11} &E_{12}\\
		E_{21}&E_{22}\end{array}
		\right) 
		$ where $E_{11}, E_{12}=E_{21}^\dagger, E_{22}$ are $5\times 5, 5\times 6,$ and $6\times 6$ matrices respectively. Following similar proof with that in Ref. \cite{Yus15}, we can prove that $E_{11}=\alpha_1 \sum_{i=0}^4|i\rangle\langle i|$ and $E_{22}=\alpha_2 \sum_{j=5}^{10}|j\rangle\langle j|$ for some $\alpha_1,\alpha_2\geq0$ by using the orthogonality relations of Eq. \eqref{eq:OR} from the sets $\{|\tilde{\psi}_i\rangle\}_{i=1}^8\cup \{|\tilde{S}\rangle\}$ and $\{|\tilde{\phi}_j\rangle\}_{j=1}^{10}\cup \{|\tilde{S}\rangle\}$ respectively. In the following, we show that $E_{12}$ is a zero matrix. Note that  the $B_2$ part of $|\tilde{\psi}_i\rangle$ and   $|\tilde{\phi}_j\rangle$ are pairwise nonorthogonal for $i=1,2,3,4$ and $j=1,2,3,4,5.$ Substitute $|\Theta_1\rangle$ and $ |\Theta_2\rangle$ in Eq. \eqref{eq:OR} by  $|\tilde{\psi}_i\rangle$ and  $|\tilde{\phi}_j\rangle$ respectively, we obtain
		$$ a_{i,j+5}=\langle i|E|j+5\rangle=0$$
		for $i=1,2,3,4$ and $j=1,2,3,4,5.$ Now  substitute $|\Theta_1\rangle$ and $ |\Theta_2\rangle$ in Eq. \eqref{eq:OR} by  $|\tilde{\psi}_{i+4}\rangle$ and  $|\tilde{\phi}_{i}\rangle$ ($i=1,2,3,4$), we obtain 
$$\begin{array}{cccc}
a_{0,6}=a_{4, 6}=0,& a_{0,7}=a_{1, 7}=0,&
a_{0,8}=a_{2, 8}=0,& a_{0,9}=a_{3, 9}=0.
\end{array}
$$
Now  substitute $|\Theta_1\rangle$ and $ |\Theta_2\rangle$ in Eq. \eqref{eq:OR} by  $|\tilde{\psi}_{i}\rangle$ and  $|\tilde{\phi}_{i+1}\rangle$ ($i=5,6$), we obtain 
$ a_{0,5}-a_{0,10}+a_{4,5}-a_{4,10}=0, $ and $ a_{0,5}-a_{0,6}+a_{1,5}-a_{1,6}=0.$ By the previous result, we get $a_{0,5}=a_{0,6}=0.$ Therefore, the matrix $E_{12}$ is a zero matrix. Hence $E$ is a diagonal matrix. At last, consider that $\langle \tilde{\mathcal{M}}| E\otimes \mathbb{I}_{B_2}| \tilde{S}\rangle=0,$ we have  $a_{1,1}=a_{6,6}$. That is, $\alpha_1=\alpha_2.$ So we conclude that $E\propto\mathbb{I}_A.$ \qed

	\end{document}